\documentclass[twocolumn,showpacs,preprintnumbers,amsmath,amssymb]{revtex4}

\usepackage{graphicx}
\usepackage{dcolumn}
\usepackage{bm}

\begin{document}

\title{Power Law Distribution of Wealth in a Money-Based
Model}
\author{Yan-Bo Xie, Bo Hu, Tao Zhou}
\author{Bing-Hong Wang}
 \email{bhwang@ustc.edu.cn}
\affiliation{%
Department of Modern Physics and The Nonlinear Science Center,\\
University of Science and Technology of China,\\
Heifei Anhui, 230026, PR China
}%

\date{\today}

\begin{abstract}
A money-based model for the power law distribution (PLD) of wealth
in an economically interacting population is introduced. The basic
feature of our model is concentrating on the capital movements and
avoiding the complexity of micro behaviors of individuals. It is
proposed as an extension of the Equ\'{\i}luz and Zimmermann's (EZ)
model for crowding and information transmission in financial
markets. Still, we must emphasize that in EZ model the PLD without
exponential correction is obtained only for a particular
parameter, while our pattern will give it within a wide range. The
Zipf exponent depends on the parameters in a nontrivial way and is
exactly calculated in this paper.
\end{abstract}


\maketitle

\section{\label{sec:level1}Introduction}

Many real life distributions, including wealth allocation in
individuals, sizes of human settlements, website popularity, words
ranked by frequency in a random corpus of text, observe the Zipf
law. Empirical evidence of the Zipf distribution of wealth [1-9]
has recently attracted a lot of interest of economists and
physicists. To understand the micro mechanism of this challenging
problem, various models have been proposed. One type of them is
based on the so-called multiplicative random process[10-21]. In
this approach, individual wealth is multiplicatively updated by a
random and independent factor. A very nice power law is given,
however, this approach essentially does not contain interactions
among individuals, which are responsible for the economic
structure and aggregate behavior. Another pattern takes into
account the interaction between two individuals that results in a
redistribution of their assets[22-25]. Unfortunately, some
attempts only give Boltzmann-Gibbs distribution of assets[24,25],
while some others[23], though exhibiting Zipf distributions, fail
to provide a stationary state.

In this paper, we shall introduce a new perspective to understand
this problem. Our model is based on the following observations:
(i) In order to minimize costs and maximize profits, two
corporations/economic entities may combine into one. This
phenomenon occurs frequently in real economic world. Simply fixing
attention on capital movements, we can equally say that two
capitals combine into one.(ii) The disassociation of an economic
entity into many small sections or individuals is also
commonplace. The bankruptcy of a corporation, for instance, can be
effectively classified into this category. Allocating a fraction
of assets for the employee's salary, a company also serves as a
good example for the fragmentation of capitals. Under some
appropriate assumptions, we shall establish a money-based model
which is essentially an extension of the Egu\'{\i}luz and
Zimmermann's (EZ) model for crowding and information transmission
in financial markets\cite{ez1,ez2}. The size of a cluster there is
now identified as the wealth of an agent here. However, analytical
results will show that our model is quite different from EZ's
\cite{ez2}, which gives PLD with an exponential cut-off that
vanishes only for a particular parameter. Here, a Zipf
distribution of wealth is obtained within a wide range of
parameters, and surprisingly, without exponential correction. The
Zipf exponent can be analytically calculated and is found to have
a nontrivial dependence on our model parameters.

This paper is organized as follows:  In section 2, the model is
described and the corresponding master equation is provided
directly. In section 3, we shall present our analytical
calculation of the Zipf exponent. Next, we give numerical studies
for the master equation, which are in excellent agreement with
analytic results. In section 5, the relevance of our model to the
real world are discussed.

\section{the model}

The money-based model contains $N$ units of money, where $N$ is
fixed. Though in real economic environment the total wealth is
quite possible to fluctuate, our assumption is not oversimplified
but reasonable, given that the production and consumption
processes are simultaneous and the resource is finite. The $N$
units of money are then allocated to $M$ agents (or say, economic
entities), where $M$ is changeable with the passage of time. For
simplicity, we may choose the initial state containing just $N$
agents, each with one unit of capital. The state of system is
mainly described by $n_s$, which denotes the number of agents with
$s$ units of money. The evolution of the system is under following
rules: At each time step, a unit of money, instead of an agent, is
selected at random. Notice that our model is much more
concentrating on the capital movement among agents rather than the
agents themselves. With probability $a\gamma/s$, the agent who
owns this unit of money is disassociated, here $s$ is the amount
of capitals owned by this agent and $\gamma$ is a constant which
implies the relative magnitude of dissociative possibility at a
macro level. After disassociation, this $s$ units of money are
redistributed to $s$ new agents, each with just one unit. It must
be illuminated that an real economic entity in most cases does not
separate in such an equally minimal way. However, with a point of
statistical view and considering analytical facility, this
simplified hypothesis is acceptable for original study. Now,
continue with our evolution rules. With probability
$a(1-\gamma/s)$, nothing is done. And with probability $1-a$,
another unit of money is selected randomly from the wealth pool.
If these two units are occupied by different agents, then the two
agents with all their money combine into one; otherwise, nothing
occurs. Thus, $1-a$ in our model is a factor reflecting the
possibility for incorporation at a macro level.

One may find that as $a$ is close to 1 and $\gamma$ is not too
small, a financial oligarch is almost forbidden to emerge in the
evolution of the system; but, if the initial state contains any
figure such as Henry Ford or Bill Gates, he is preferentially
protected. Note that the bankruptcy probability of moneybags is
inverse proportional to their wealth ranks, and the possibility of
being chosen is proportional to $sn_s/N$; thus, the Doomsday of a
tycoon comes with possibility $an_s\gamma/N$, which is extremely
small for large $s$. Meanwhile, the vast majority, if initially
poor, is perpetually in poverty, with no chance to raise the
economic status any way. In addition, if middle class exists at
first, it will not disappear or expand in the foreseeable future.
Again, it may be interesting to argue that when $a$ is slightly
above zero, the merger process is prevailing and overwhelming, and
all the capitals are inclined to converge. In this case, though
the rich are preferentially protected, the trend in the long run
is to annihilate them until the last. Of course, one-agent game is
trivial. Likewise, it is not appealing to observe the system when
$\gamma$ goes to 0 and $a$ to 1, since both merger and
disintegration are nearly impossible--in other words, all the
capitals are locked, thus the wealth pool is dead at any time.

Following Refs.\cite{ez2,x1,x2} in the case of $N\gg 1$, we give
the master equation for $n_s$
\begin{equation}
{\partial n_s\over\partial t}={1-a\over N}\sum_{r=1}^{s-1} rn_r(s-r)n_{s-r}
-2(1-a)sn_s-asn_s{\gamma\over s}
\end{equation}
for $s>1$ and
\begin{eqnarray}
{\partial n_1\over\partial t}
&&=-2(1-a)n_1+a\sum_{s=2}^{\infty}s^2n_s{\gamma
\over s}\nonumber\\
&&=-2(1-a)n_1+a\gamma(N-n_1)
\end{eqnarray}
where the identity
\begin{equation}
\sum_{s=1}^{\infty} sn_s=N
\end{equation}
has been used. We must point out that Eq.(1) is almost the same as
the master equation derived in Ref.\cite{ez2} for the EZ model
except for an additional factor $\gamma/s$ in the third term on
the right hand side of Eq.(1). Notice that this term is
significant because otherwise the frequency of the disintegration
for large $s$ agents would be too high.

Now we introduce $h_s=sn_s/N$, which indicates the ratio of wealth
occupied by agents in rank $s$ to the total wealth, and
$\alpha=a\gamma/2(1-a)$, that represents the maximum ratio of the
disintegration possibility to the merger probability in the whole
economic environment. Then, one can give the equations for the
stationary state in a terse form:
\begin{equation}
h_s={s\over 2(s+\alpha)}\sum_{r=1}^{s-1}h_rh_{s-r}
\end{equation}
and
\begin{equation}
h_1={\alpha\over 1+\alpha}
\end{equation}
According to the definition of $h_s$, it should satisfy the
normalization condition Eq.(3)
\begin{equation}
\sum_{s=1}^{\infty}h_s=1
\end{equation}
When $\alpha$ is less than a critical value $\alpha_c=4$ which
will be determined numerically in section 4, one can show that
Eqs.(4-5) does not satisfy the normalization condition Eq.(3).
This inconsistency implies that when $\alpha<\alpha_c$ the state
with one agent who has all the $N$ units of money becomes
important\cite{x1,x2}. In this case, the finite-size effect and
the fluctuation effect become nontrivial and the master equations
(1-3) is no longer applicable to describe the system\cite{x1,x2}.
In this paper, we shall restrict our discussion to the case
$\alpha>\alpha_c$.

\section{Analytic results}

When $\alpha>\alpha_c$, one can show that $h_s\to A/s^{\eta}$ for
sufficiently large $s$ with
\begin{equation}
\eta={\alpha\over\sum_{r=1}^{\infty} rh_r}
\end{equation}
Notice that this equation is only consistent when $\eta>2$ because
otherwise the sum $\sum_{r=1}^{\infty}rh_r$ would be divergent,
and thus $h_s\to A/s^{\eta}$ becomes an inconsistent formula.

The derivation of Eq.(7) is described as follows: When $s$ is sufficiently
large
\begin{eqnarray}
h_s&&={s\over 2(s+\alpha)}\sum_{r=1}^{s-1} h_rh_{s-r}\nonumber\\
&&\approx {s\over s+\alpha}(\sum_{r=1}^{s^\delta} h_{s-r}h_r
+h_s O({1\over s^{2\delta\eta-1-\eta}}))\nonumber\\
&&\approx {s\over s+\alpha}\sum_{r=1}^{s^\delta}(h_s-{dh_s\over ds}r)h_r
\nonumber\\
&&\approx (1-{\alpha\over s})[h_s\sum_{r=1}^\infty h_r-{dh_s\over ds}\sum_{r=1}
^{\infty} rh_r]+h_sO({1\over s^{\delta(\eta-1)}})\nonumber\\
&&\approx (1-{\alpha\over s})[h_s-{dh_s\over ds}\sum_{r=1}^{\infty} rh_r]
\end{eqnarray}
where $\delta<1$ but is close to 1, $\delta(\eta-1)>1$ and
$2\delta\eta-1-\eta>1$.  Therefore
$${dh_s\over ds}=-{h_s\over s}{\alpha\over\sum_{r=1}^{\infty} rh_r}$$
which gives that as $s\to\infty$
\begin{equation}
h_s={A\over s^{\eta}}
\end{equation}

The value of $\sum_{r=1}^{\infty}rh_r$ can be further evaluated:

Introducing the generating function
\begin{equation}
G(x)=\sum_{r=1}^{\infty} x^r h_r
\end{equation}
one can rewrite Eq.(4) as
$$x(G'-h_1)+\alpha (G-h_1x)=xG'+\alpha(G-x)=xG'G$$
or
\begin{equation}
G'x(G-1)=\alpha (G-x)
\end{equation}
with the initial condition
\begin{equation}
G(0)=0
\end{equation}
Since $h_s\to A/s^\eta$ as $s\to\infty$, $G$ is only defined in
the interval $|x|\leq 1$.  From Eq.(6), we also have $G(1)=1$.
What we need to calculate is just
$$G'(1)=\sum_{r=1}^{\infty}rh_r$$
Since the left and the right hand sides of Eq.(11) are both zero at $x=1$, we
differentiate both sides by $x$ and obtain
$$G''x(1-G)+G'(1-G)-xG'^2=\alpha (1-G')$$
Let $x\to 1$ and one finds that $G''(1-G)$ vanishes in this limit
provided $\eta>2$, thus
\begin{equation}
G'^2(1)-\alpha G'(1)+\alpha=0
\end{equation}
One immediately obtains that
\begin{equation}
\sum_{r=1}^{\infty} rh_r={\alpha-\sqrt{\alpha^2-4\alpha}\over 2}
\end{equation}
and the exponent
\begin{equation}
\eta={2\over 1-\sqrt{1-4/\alpha}}
\end{equation}
which is a positive real number for $\alpha\geq 4$. Notice that
when $\alpha=4$, the exponent $\eta=2$. This implies that our
calculation is self-consistent, provided Eq.(6). In sum, we find
from the master equation that $h_s$ obeys PLD when $s$ is
sufficiently large and $\alpha>4$. It may be important to point
out that when $s$ is small, $h_s$ also approximately obeys the
PLD, and the restriction $\alpha>4$, introduced for the sake of
discussing master equation, can be actually relaxed. This argument
has been tested by the simulator investigation, which supplies the
gap of analytical tools and verifies the analytical outcome.

\section{Numerical results}

\begin{table}
\caption{\label{tab:table1} The results of $H$ for various value
of $\alpha$. }
\begin{tabular}
{|c|c|}\hline $\alpha$ & $H$  \\ \hline 3.0 & 0.9940886  \\
\hline 3.5 & 0.9997818  \\ \hline 3.6 & 0.9999214  \\ \hline 3.7 &
0.9999743
\\ \hline 3.8 & 0.9999922  \\ \hline 3.9 & 0.9999977  \\ \hline
4.0 & 0.9999995  \\ \hline 4.1 & 0.9999999  \\ \hline 4.2 &
1.0000000  \\ \hline 4.3 & 1.0000000  \\ \hline 4.4 & 1.0000000
\\ \hline 4.5 & 1.0000000  \\ \hline 5.0 & 1.0000000  \\ \hline
6.0 & 1.0000000  \\ \hline
\end{tabular}
\end{table}

We have numerically calculated the number
$$H=\sum_{r=1}^{\infty} h_r$$
based on the recursion formula Eq.(4) with the initial condition
Eq.(5).  Table.1 lists the results of $H$ for various value of
$\alpha$.  From Table.1, one immediately find that the
normalization condition is satisfied for $\alpha>\alpha_c=4$,
which, again, indicates consistency of related equations.

Fig.1-2 show $h_s$ as a function of $s$ in a log-log scale for
$\alpha=10$, $\alpha=4.5$, respectively.  From Fig.1, one can see
that $h_s$ conforms to PLD for $s\gg1$ with the exponent $\eta$
given by Eq.(15). Fig.2 indicates that $h_s$ observes the Zipf law
for nearly all $s$ with $\eta=3.0$.

The fitted exponents for various values of $\alpha$ are plotted in
Fig.3. They are given by
$${\ln (h_{900}/h_{1000})\over \ln (1000/900)}$$  Fig.3 also exhibits the analytic
results from Eq.(15). The analytic outcome fits the exponents
calculated from recursion quite well for $\alpha>4.2$. However,
when $\alpha\to 4.0$, discrepancy is obvious, since the
convergence of $h_s$ to the correct power law is then very slow.

We have also performed computer simulation, which gives excellent
agreement with theoretical results derived from Eqs.(4-5) for
$\alpha=8$ and $s\leq 10$, see Fig.4. For more about our simulator
investigation and further analysis for $\alpha<4$, see Ref.
\cite{hu}.

\begin{figure}
\scalebox{0.8}[0.75]{\includegraphics{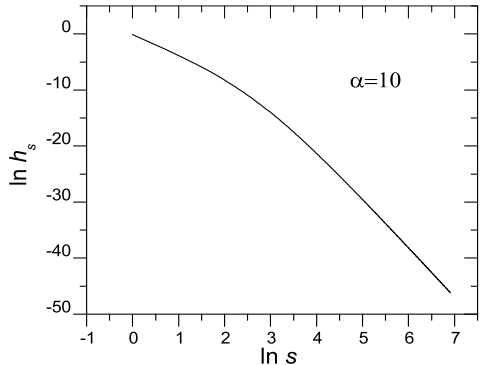}}
\caption{\label{fig:epsart} The dependence of $h_s$ on $s$ in a
log-log scale for $\alpha=10$.}
\end{figure}

\begin{figure}
\scalebox{0.8}[0.75]{\includegraphics{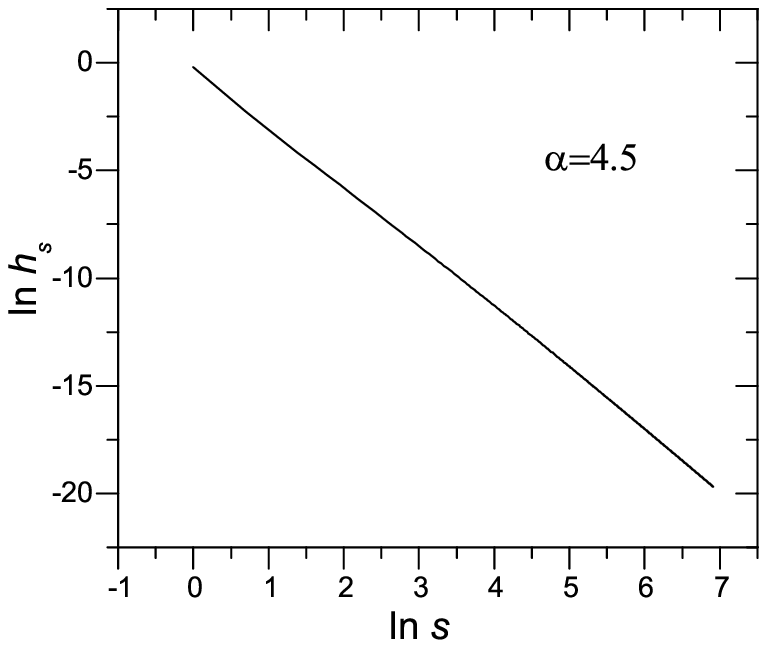}}
\caption{\label{fig:epsart} The dependence of $h_s$ on $s$ in a
log-log scale for $\alpha=4.5$.}
\end{figure}

\begin{figure}
\scalebox{0.8}[0.75]{\includegraphics{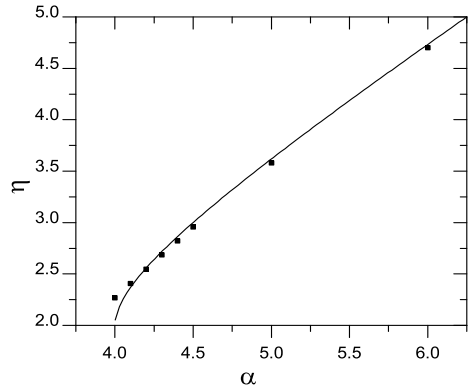}}
\caption{\label{fig:epsart} The calculated exponent $\eta$ for
different values of $\alpha$. Black squares represent the
numerical results of $\eta$ obtained from $h_s$ using the
extropolation method, see text.  The solid line represents the
analytic result Eq.(15).}
\end{figure}

\begin{figure}
\scalebox{0.8}[0.75]{\includegraphics{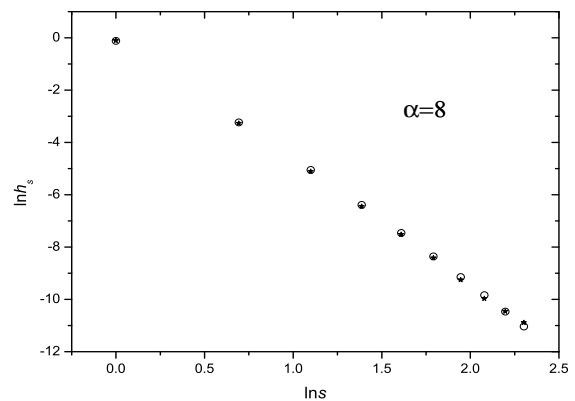}}
\caption{\label{fig:epsart} $h_s$ for $\alpha=8$ from both
numerical calculation and computer simulation. Black stars
represent outcome of computer simulation for $N=2.5\times 10^{5}$,
$\gamma=2$ and $a=0.88889$. Total $2\times 10^6$ time steps were
run and the final $5\times 10^5$ time steps were used to count
$n_s$ statistically. The circles represent the theoretical results
derived from Eqs.(4-5).}
\end{figure}

\section{Discussions}

In this paper, we have introduced a so-called money-based model to
mimic and study the wealth allocation process. We find for a wide
range of parameters, the wealth distribution $n_s\sim
A/s^{\eta+1}$ with $\eta$ given by Eq.(15) for sufficiently large
$s$. The crucial difference between our model and the EZ model is
that the dissociative probability $\Gamma_d$ of an economic
entity, after he/she is picked up, is proportional to $1/s$ in our
model. However, the corresponding probability in the EZ model is
simply proportional to 1. This difference gives rise to divergent
behaviors of $n_s$. In the EZ model, $n_s\sim A/s^{2.5}
\exp(-\alpha s)$ for large $s$ \cite{ez2}. When $n_s$ is
interpreted as the number of individuals who own $s$ units of
assets, the choice of $\Gamma_d\sim O(1/s)$ is reasonable.
Actually, since at the first step, we randomly picked up a unit of
money, the individual who owns $s$ units of assets is picked up
with a probability proportional to $s$. According to the
observation in real economic life, large companies or rich men are
often much more robust than small or poor ones when confronting
economic impact and fierce competition. If $\Gamma_d\sim O(1)$,
the overall dissociation frequency would be proportional to $s$
which is totally unreasonable.

In real economic environment, capitals and agents behave similarly
at some point. For instance, they both ceaselessly display
integration and disintegration, driven by the motivation to
maximize profits and efficiency. This mechanism updates the system
every time, and gives rise to clusters and herd behaviors.
Furthermore, in an agent-based model, it is usually indispensable
to consider the individual diversity that is all too often hard to
deal with. When it comes to the money-based model, this micro
complexity may be considerably simplified. Finally, the conceptual
movement and interaction among capitals is not as restricted by
space and time as between agents. Therefore, when econophysics is
much more interested in the behaviors of capitals than that of
agents, it is recommendable to adopt such a money-based model.

The methodology to fix our attention on the capital movements,
instead of interactions among individuals, will bring a lot of
facility for analysis; moreover, using such random variables as
$\gamma$ and $a$ to represent the macro level of the micro
mechanism also help us find a possible bridge between the
evolution of the system and the protean behaviors of individuals.
Whether the bridge is steady or not can only be tested by further
investigation.

\begin{acknowledgments}
This work has been partially supported by the State Key
Development Programme of Basic Research (973 Project) of China,
the National Natural Science Foundation of China under Grant
No.70271070 and the Specialized Research Fund for the Doctoral
Program of Higher Education (SRFDP No.20020358009)
\end{acknowledgments}

\end{document}